\documentclass[conference]{IEEEtran}

\usepackage{booktabs}
\usepackage{multirow}
\usepackage{array}
\usepackage{tabularx}
\usepackage[table]{xcolor}

\definecolor{tableheader}{RGB}{238,242,247}
\definecolor{tableavg}{RGB}{229,238,248}

\newcolumntype{Y}{>{\centering\arraybackslash}X}

\ifCLASSINFOpdf
\else
\fi
\usepackage{amsmath}
\usepackage{amssymb}
\usepackage{amsfonts}
\usepackage{bm}
\usepackage{mathtools}
\usepackage{graphicx}
\usepackage{booktabs}
\usepackage{array}
\usepackage{multirow}
\usepackage{makecell}
\usepackage{adjustbox}

\usepackage{xcolor}
\usepackage{url}
\usepackage{cite}

\begin{document}
\title{What's Your NIC Whispering?\\Network Threat Behavior Recognition
via NIC Electromagnetic Side-Channel Leakage}

\author{
\IEEEauthorblockN{Hongchao Wang\IEEEauthorrefmark{1},
Linrui Li\IEEEauthorrefmark{1},
Yunkai Zou\IEEEauthorrefmark{2},
Zhenduo Hou\IEEEauthorrefmark{2},
Yilin Zhang\IEEEauthorrefmark{1},
Haoyang Pu\IEEEauthorrefmark{1},
Wen Chen\IEEEauthorrefmark{1},
Jierui Chen\IEEEauthorrefmark{3}}

\IEEEauthorblockA{\IEEEauthorrefmark{1} Sichuan University\\
wanghc@stu.scu.edu.cn, rLarry@stu.scu.edu.cn, zhangyilin3@stu.scu.edu.cn, puhaoyang78@stu.scu.edu.cn, wenchen@scu.edu.cn}

\IEEEauthorblockA{\IEEEauthorrefmark{2} Nankai University\\
zouyunkai@nankai.edu.cn, joehou13@nankai.edu.cn}

\IEEEauthorblockA{\IEEEauthorrefmark{3} Institute of Information Engineering, Chinese Academy of Sciences\\
jeffy353866@gmail.com}
}

\maketitle

\begin{abstract}

Conventional network threat detection primarily relies on packet-level, flow-level, or host-level telemetry. This paper investigates a different observation surface: unintended electromagnetic (EM) emissions generated by network interface card (NIC) activity, and asks whether such physical leakage contains sufficiently structured information for network threat-behavior recognition. We present NICWhisper, which externally captures NIC EM emissions, transforms raw measurements into time–frequency representations, and recognizes network behaviors without inspecting packet contents or host-side runtime states. Rather than competing with traffic-based detection, NICWhisper exploits the physical manifestation of traffic-driven NIC activity, whose timing, rate, concurrency, and burst organization naturally shape the measured EM leakage. We construct a NIC EM dataset covering active benign workloads and seven representative threat behaviors under diverse execution conditions, and systematically evaluate signal dependence, execution variation, measurement perturbation, and cross-device transfer. NICWhisper achieves 80.67\% Macro-F1 across eight behavior classes, while further experiments show that the observed behavior-related information extends beyond simple signal magnitude and remains partially transferable across execution conditions and NIC hardware. These results establish NIC EM leakage as a complementary physical observation source for network security monitoring when direct access to conventional traffic or host telemetry is limited or undesirable.

\end{abstract}


% \fi
\IEEEpeerreviewmaketitle

\section{Introduction}
\label{sec:introduction}

Network intrusion detection and defense systems are a core component of modern computing security.
Existing systems mainly rely on traffic inspection, host logs, and
endpoint monitoring agents~\cite{alkasassbeh2023intrusion,khraisat2019survey,han2024ecnet,zipperle2022provenance,hids_survey} to identify malicious network activities
and provide rich security context. However, these defensive components
are increasingly targeted by attackers and may themselves become
potential entry points or objects of evasion~\cite{uetz2024siem,alachkar2025eviledr}. Moreover, on some closed,
mission-critical, legacy, or resource-constrained devices, deploying
continuous traffic inspection or host-side monitoring may consume
additional computing, memory, and I/O resources and consequently affect
normal system operation. These practical limitations motivate the
exploration of an observation source that is independent of the
monitored software stack. In this work, we investigate whether the
electromagnetic side channel of the network interface card (NIC) can
serve as a complementary mechanism for network threat detection by
capturing the unintended EM emissions produced during NIC operation.

\begin{figure}
    \centering
    \includegraphics[width=0.96\linewidth]{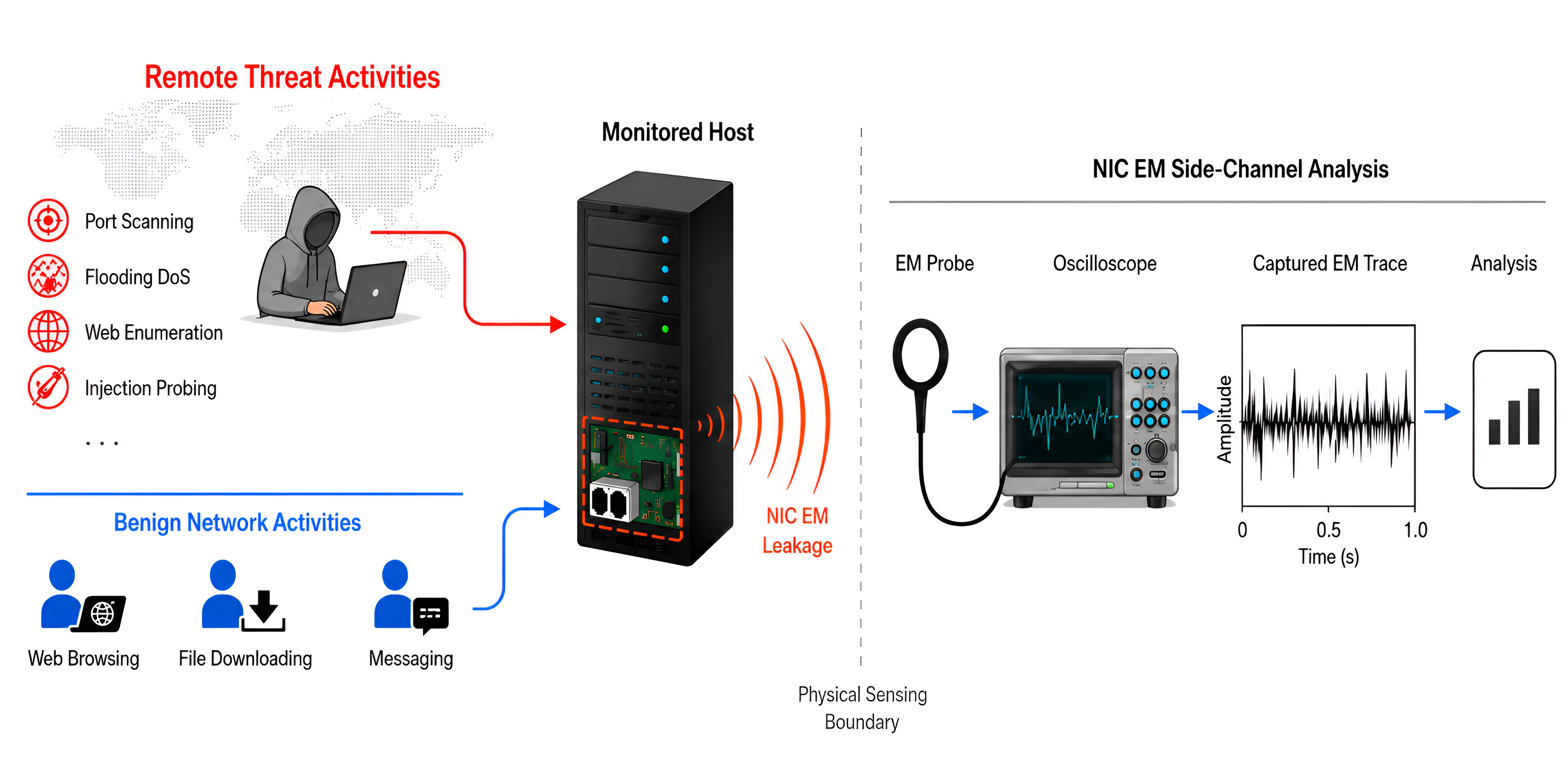}
    \caption{Overview of NICWhisper for network-behavior recognition using externally captured NIC electromagnetic emissions.}
    \label{fig:intro}
\end{figure}

As illustrated in Fig.~\ref{fig:intro}, a probe placed near the target
NIC captures its unintended EM emissions, which are analyzed on an
independent acquisition host. NICWhisper uses only this physical
observation and does not inspect packets or host-side runtime states.

Although the NIC provides an intuitive physical interface between network activities and electromagnetic emissions, whether such leakage contains sufficient information for distinguishing different network behaviors remains unclear. Unlike conventional network telemetry, where packet contents, flow statistics, and protocol semantics are directly available, NIC electromagnetic leakage provides an indirect physical observation of NIC processing activity. Its characteristics are naturally shaped by how a network behavior is executed, including request timing, rate, concurrency, and burst organization. These traffic-induced dynamics are therefore part of the physical manifestation of network behavior at the NIC. More broadly, prior work has shown that network activity can also
be exposed through indirect timing and transport-layer side channels
~\cite{gast2024snailload,yu2024athena}.
Accordingly, the first question we investigate is whether different network behaviors induce sufficiently distinguishable physical patterns in NIC electromagnetic emissions.

\begin{figure}[t]
    \centering
    \includegraphics[width=\linewidth]{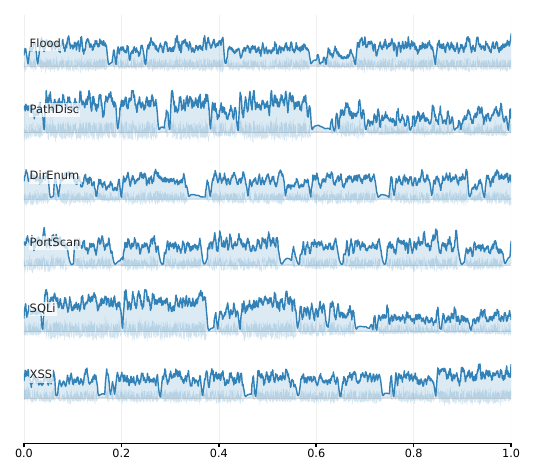}
    \caption{Representative NIC electromagnetic traces collected under different network behaviors. The measurements exhibit observable variations in burst organization, activity duration, and temporal dynamics, while retaining substantial overlap.}
    \label{fig:waveform}
\end{figure}

Figure~\ref{fig:waveform} shows representative EM observations
under different network activities. The traces exhibit partially
overlapping yet behavior-dependent temporal variations, motivating
the systematic analysis that follows.

However, exploiting NIC electromagnetic leakage as a security observation source introduces two fundamental challenges. First, the measured EM signal reflects the physical consequences of NIC activity rather than explicit network semantics, making it necessary to determine whether different network behaviors induce distinguishable physical structures. Second, such structures may vary with tools, execution configurations, background activities, and hardware conditions, raising the question of whether behavior-related information persists beyond a specific experimental realization. 
Motivated by these observations, we introduce NICWhisper, a systematic study of NIC electromagnetic leakage as a complementary physical observation channel for network behavior recognition. NICWhisper aims to characterize whether such physical patterns provide stable and discriminative evidence across different network behaviors and execution conditions. To this end, NICWhisper employs a simple time–frequency analysis pipeline that converts raw electromagnetic measurements into discriminative representations and evaluates their capability under diverse network scenarios.

To evaluate the feasibility of this new observation channel, we
construct a NIC electromagnetic dataset containing active
benign workloads and seven representative network threat behaviors.
The dataset includes variations in network tools, execution
configurations, target conditions, and concurrent benign activities,
which allows us to examine whether the observed electromagnetic
patterns are associated with general network behaviors rather than
fixed experimental artifacts. Extensive experiments demonstrate that
NIC electromagnetic leakage preserves discriminative behavior-related
information and can serve as a complementary observation source for
network security analysis.

Our contributions are summarized as follows:

\begin{itemize}

\item We release a NIC electromagnetic leakage dataset for
network behavior recognition, covering active benign workloads and
seven representative threat behaviors under varied tools, execution
configurations, and concurrent network activities.

\item We present, to the best of our knowledge, the first systematic
study of NIC electromagnetic leakage as a physical observation channel
for network behavior recognition, and characterize how aggregate NIC
activity exposes behavior-related signal structures.

\item We develop a simple time--frequency analysis and recognition pipeline
and conduct comprehensive experiments to characterize its effectiveness,
signal dependence, sensitivity, execution variation, and cross-device
transfer.

\end{itemize}

\section{Motivation and Threat Model}
\label{sec:motivation_threat}

\subsection{Motivation of NICWhisper}
\label{subsec:motivation}

Network traffic inspection and host monitoring provide rich semantic information and remain the primary means of network threat detection. NICWhisper complements these mechanisms with an external physical observation channel acquired independently of the monitored software stack.

Such an observation can be useful for closed, legacy,
mission-critical, or resource-constrained systems where modifying the
software stack or continuously running monitoring agents is undesirable
or impractical. It may also provide security evidence acquired outside
the software environment being examined.

The NIC is a natural observation point because network communication
must eventually be processed by the network interface.  Recent work has likewise explored communication
controllers as strategic observation points for security monitoring
~\cite{cayre2024oasis}. Packet
transmission and reception trigger frame processing, buffering,
controller operations, host-interface communication, and physical-layer
activity. The corresponding electrical activity may produce
behavior-dependent electromagnetic emissions.

NICWhisper captures these emissions using an external probe and analyzes
them on an independent host. It does not inspect packet contents or
require software instrumentation of the monitored system. The intended
deployment targets a limited number of fixed and security-sensitive
devices for which an additional physical observation is valuable.

\subsection{Threat Model}
\label{subsec:threat_model}

\textbf{Monitored setting.}
We consider a server or networked device whose evaluated communication
passes through a designated NIC. The defender owns or manages the
device and is authorized to place an EM probe near the NIC. The sensing
equipment and analysis host are trusted, and their configuration remains
fixed during recognition.

NICWhisper receives only the captured EM trace as input. Packet
payloads, flow records, host logs, system calls, and process states are
not available to the recognizer.

\textbf{Network activities.}
A remote adversary communicates with the monitored system through the
target NIC and may perform flooding, port scanning, service probing,
web enumeration, SQL-injection probing, or cross-site-scripting
probing. The attacker may vary tools, request rates, concurrency,
timing, targets, and attack inputs.

Meanwhile, benign users may browse web pages, request content, or
download files from the same server. Threat and benign activities may
therefore coexist on the monitored NIC. For a threat-related observation, we represent the acquired trace
abstractly as

\begin{equation}
    \mathbf{x}
    =
    \mathcal{H}_{h,p}
    \left(
        \mathcal{A}_{y,\kappa}(\bm{\theta})
        \oplus
        \mathcal{B}(\bm{\phi})
    \right)
    +
    \mathbf{n},
    \label{eq:em_observation}
\end{equation}

where \(\mathcal{A}_{y,\kappa}(\bm{\theta})\) denotes threat behavior
\(y\) generated by tool \(\kappa\) under configuration
\(\bm{\theta}\), and \(\mathcal{B}(\bm{\phi})\) denotes concurrent
benign activity. The mapping \(\mathcal{H}_{h,p}\) captures the
hardware- and sensing-dependent transformation from the composite
network workload to the measured EM trace.

This abstraction does not attribute the leakage to a specific internal
circuit. The captured signal may contain the combined activity of the
NIC controller, physical-layer interface, host interface, and local
supporting circuitry.

\textbf{Recognition scope.}
NICWhisper focuses on representative network activities commonly observed during the early discovery, probing, and exploitation stages of remote attacks. The evaluated behaviors include flooding, port scanning, service probing, directory enumeration, web-path discovery, SQL-injection probing, and cross-site-scripting probing. Active benign access is included as a separate class, covering ordinary web browsing, content requests, and file downloading.

The classes are defined by observable network behaviors rather than
complete attack campaigns or strict MITRE ATT\&CK technique labels.
NICWhisper therefore associates each EM observation with one of the
predefined behavior families based on aggregate NIC activity.

\section{Background}
\label{sec:background}

\subsection{Physical Side Channels for Security Monitoring}

Runtime electrical activity can induce externally measurable power and
electromagnetic variations. Beyond traditional side-channel leakage,
prior studies have used such physical signals for external security
monitoring, including intrusion and malware detection~\cite{khan2021idea,ding2020deeppower}.
Other work further shows that EM emissions can preserve higher-level
runtime structure, such as visual or neural-network execution
information~\cite{demeulemeester2023spectrem,long2024emeye,li2025emiris,onishi2025sound,xiao2026modelspy}.

These studies motivate physical sensing as an observation source
outside the monitored software stack. However, they do not establish
whether remotely generated network behaviors can be recognized from
aggregate NIC electromagnetic leakage, particularly in the presence
of concurrent benign traffic.

\subsection{NIC Electromagnetic Side Channel}
\label{subsec:nic_em_channel}

A network interface card (NIC) mediates communication between the host
and the external network. Packet transmission and reception activate
multiple functional components of the NIC, including the host
interface, controller/MAC logic, buffering and DMA operations, and the
physical-layer circuitry. The resulting time-varying electrical
activity can produce unintended electromagnetic (EM) emissions that are
observable outside the device~\cite{ayoub2024bluescream,ayoub2025phasesca}.

\begin{figure}[tpb]
    \centering
    \includegraphics[width=1\linewidth]{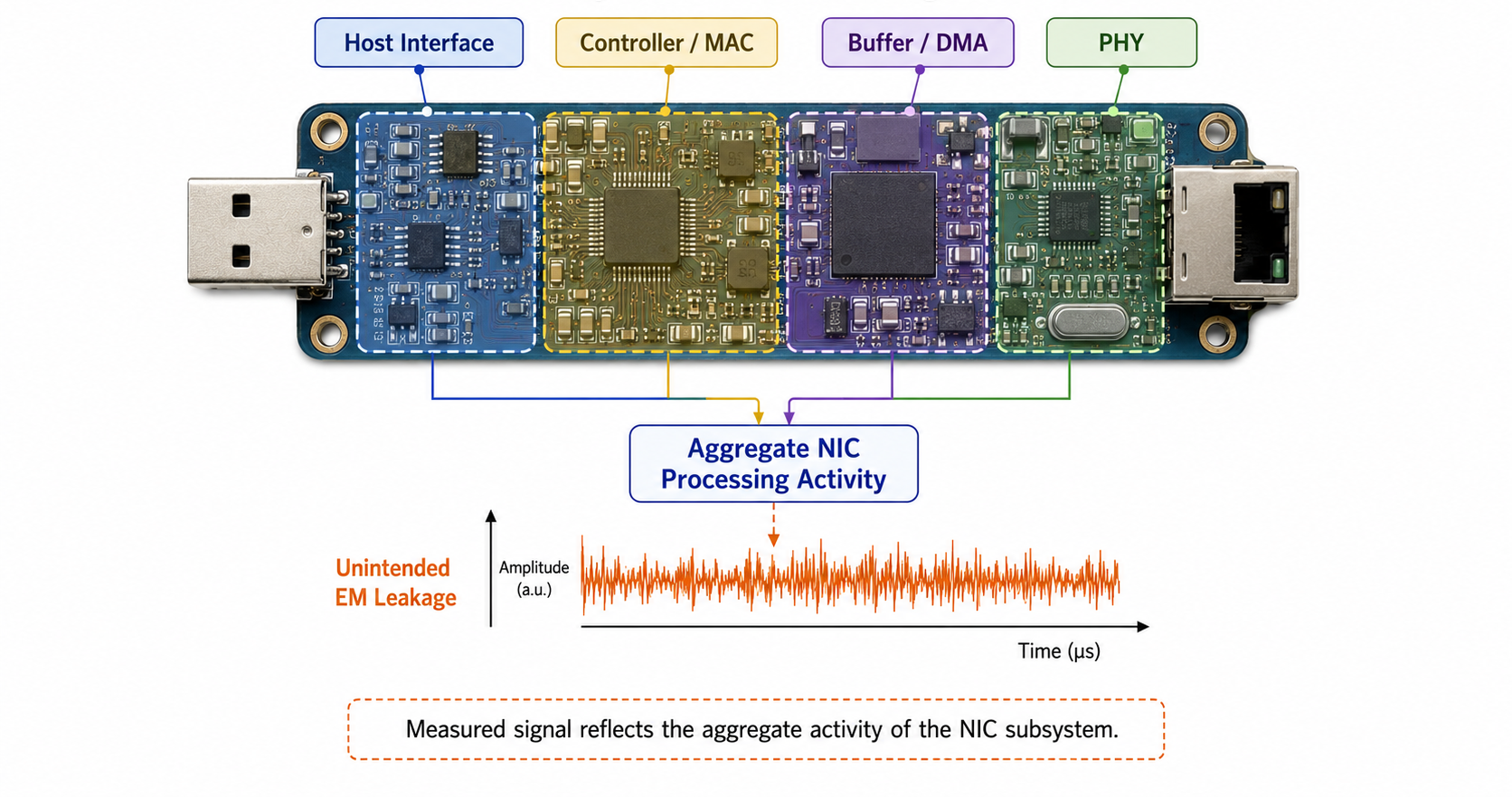}
    \caption{Conceptual relationship between NIC processing activity
    and unintended electromagnetic emissions. Multiple functional
    components contribute to aggregate NIC activity observed through
    the external EM channel.}
    \label{fig:nic_leakage_mechanism}
\end{figure}

As illustrated in Fig.~\ref{fig:nic_leakage_mechanism}, NICWhisper
treats the measured EM signal as an aggregate physical observation of
NIC processing activity. The captured leakage may contain contributions
from multiple parts of the NIC subsystem and supporting circuitry;
therefore, we do not assume that individual sources can be separately
observed or that a measured waveform can be attributed to a specific
circuit block. The highlighted functional regions illustrate the major
processing stages involved in network communication rather than a
circuit-level leakage decomposition.

Accordingly, the captured EM trace is not a direct representation of Ethernet frames, packet contents, or protocol fields. Instead, it reflects aggregate electrical activity induced by traffic processing, whose temporal and spectral organization is naturally shaped by execution characteristics such as request timing, rate, concurrency, burst structure, and transfer intensity. These characteristics are intrinsic to how network behaviors are realized at the NIC and form part of the physical mechanism through which network behavior becomes observable in the EM domain. NICWhisper therefore treats such traffic-induced physical characteristics as behavior-related evidence rather than attempting to separate them from the network behavior that produces them. Since the same high-level behavior may vary across tools, parameters, and concurrent benign traffic, the resulting EM traces are modeled as statistical behavior-related evidence rather than deterministic waveform signatures.

\subsection{Preliminary Signal Characterization}

We perform a preliminary characterization to examine whether NIC
electromagnetic measurements retain behavior-related information and
whether such information is captured by coarse whole-trace summaries.
The measurements contain active benign access and representative threat
behaviors collected using the same NIC and sensing configuration.

\begin{figure}[tpb]
    \centering
    \includegraphics[width=\linewidth]
    {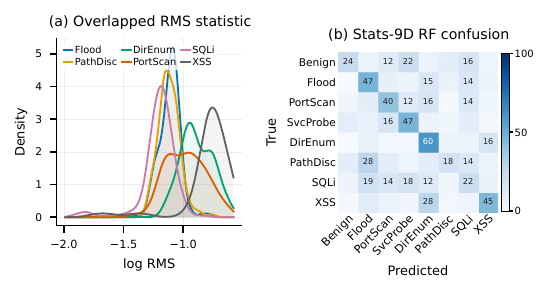}
    \caption{Preliminary characterization of NIC electromagnetic
    measurements. (a) Network behaviors exhibit different log-RMS
    tendencies, but their distributions substantially overlap.
    (b) A random forest using nine whole-trace descriptors produces a
    structured confusion matrix while retaining considerable cross-class
    confusion.}
    \label{fig:feasibility}
\end{figure}

We first calculate the root-mean-square value of each one-second trace.
Figure~\ref{fig:feasibility}(a) shows that the behavior classes exhibit
different distribution tendencies, indicating that network workloads
affect the overall measured activity. However, the distributions
substantially overlap, and a single global signal level cannot reliably
separate benign access from different threat behaviors.

We then extract nine conventional whole-trace descriptors covering
amplitude, energy, distribution shape, and coarse spectral properties,
and use them to train a random-forest classifier. As shown in
Fig.~\ref{fig:feasibility}(b), the resulting confusion matrix exhibits a
visible diagonal structure, providing initial evidence that the
measurements retain behavior-related information. Considerable confusion
nevertheless remains among benign access, scanning, enumeration, and
application-layer probing.

This diagnostic classifier evaluates how much behavior-related information is retained by conventional whole-trace summaries. The results show that they
do, but also discard important temporal organization.

A whole-trace RMS value removes temporal and spectral structure, while a
global spectrum does not preserve when particular frequency components
occur. A time--frequency representation instead retains both the
temporal evolution and spectral-energy distribution of NIC activity,
making it suitable for representing sustained processing, intermittent
communication, repeated bursts, and irregular probing sequences.

\section{NICWhisper Design}
\label{sec:design}

\subsection{Design Overview}
\label{subsec:design_overview}

NICWhisper observes the aggregate physical activity induced by network processing at the NIC and examines whether this activity retains sufficiently stable behavior-related information for recognition.

\begin{figure*}[t]
    \centering
    \includegraphics[width=\textwidth]{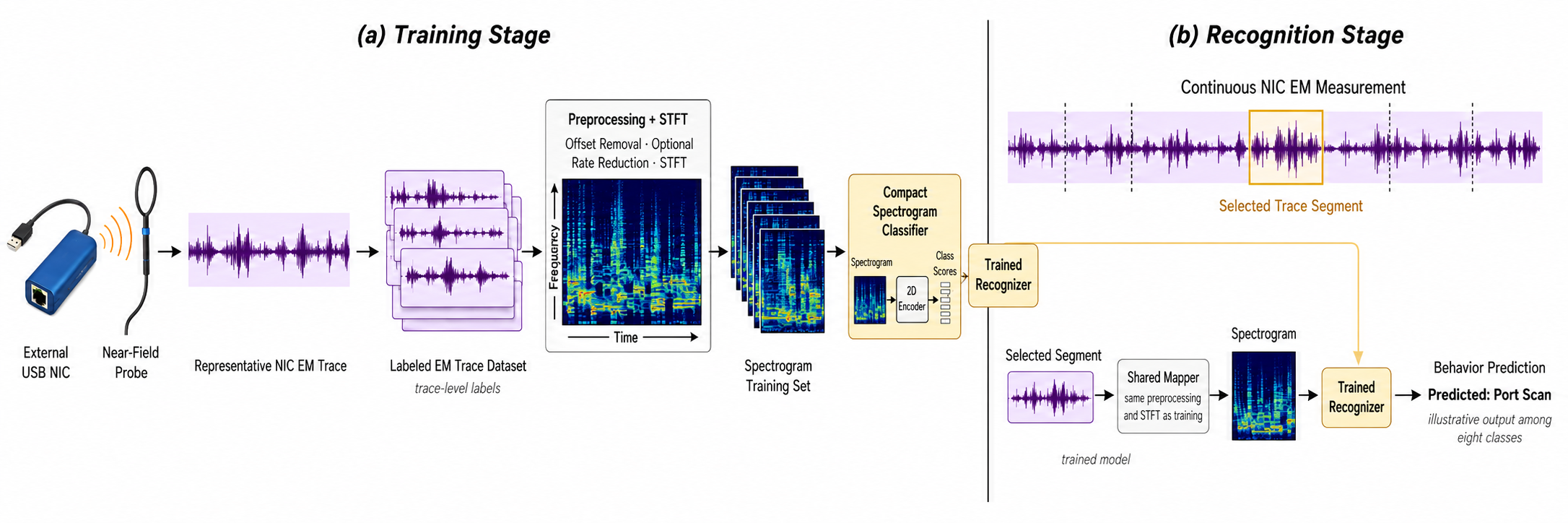}
    \caption{Overview of NICWhisper. During training, externally captured
    NIC EM traces are associated with network-behavior labels and
    transformed into time--frequency representations for classifier
    training. For a new EM observation, NICWhisper applies the same
    signal conditioning and STFT mapping before predicting the
    corresponding network behavior.}
    \label{fig:design_overview}
\end{figure*}

Figure~\ref{fig:design_overview} presents the overall design.
NICWhisper consists of two conceptual stages. During training, network
behaviors are exercised against a monitored service while a near-field
probe externally captures the electromagnetic emissions of the NIC.
The resulting traces are associated with behavior-level labels and
converted into time--frequency representations, from which a
spectrogram-based classifier is trained. For a new NIC EM observation,
the same signal conditioning and time--frequency transformation are
applied before behavior prediction. The two stages therefore share the
same physical observation and representation pipeline; their difference
lies only in whether the resulting representation is used to train or
query the classifier.

Two challenges guide the design of NICWhisper.

\textbf{Challenge 1: Execution diversity.}
The same network behavior can be realized by different tools,
parameters, timing patterns, and inputs. A dataset tied to a single
execution per class may therefore capture implementation-specific
artifacts rather than behavior-related physical characteristics.
NICWhisper introduces diverse executions during data construction while
maintaining behavior-level supervision.

\textbf{Challenge 2: Temporal--spectral organization.}
Behavior-related evidence may appear in signal intensity, burst timing,
repetition, and spectral-energy evolution. NICWhisper therefore
transforms each EM observation into a time--frequency representation
that preserves these complementary structures.

\subsection{Network Behavior and EM Trace Construction}
\label{subsec:trace_construction}

The first requirement of NICWhisper is to establish a meaningful
correspondence between a network behavior and the physical activity
observed at the NIC. We construct this relationship using a functional
web service rather than an idle network interface or an isolated
attack-only setup.

Specifically, we deploy an active Blog service as the monitored
application. Normal observations contain legitimate browsing, content
requests, and file-transfer activities with varied resources,
navigation order, and request timing. The Normal class therefore
represents active legitimate network usage rather than an idle NIC
baseline.

We consider one benign behavior together with seven representative
threat behaviors: flooding, port scanning, service probing, directory
enumeration, web-path discovery, SQL-injection probing, and
cross-site-scripting probing. Let
\(\mathcal{Y}=\{y_1,\ldots,y_C\}\) denote the behavior set. For a
behavior \(y\), we describe one concrete execution as
\begin{equation}
    a=(y,\kappa,\boldsymbol{\theta}),
    \label{eq:behavior_execution}
\end{equation}
where \(\kappa\) denotes the tool or implementation realizing the
behavior and \(\boldsymbol{\theta}\) denotes its execution
configuration. Depending on the behavior, the latter may include request
rate, concurrency, timing, scan mode, target selection, wordlist, path
set, or attack input.

\textbf{Key Design 1: Behavior-level supervision under execution
diversity.}
For each behavior \(y\), NICWhisper varies both the implementation
\(\kappa\) and behavior-specific configuration \(\theta\) during
collection while keeping the supervision target unchanged. The learning task targets behavior-level patterns shared across different tools and command configurations.

Table~\ref{tab:behavior_workloads} summarizes the behaviors and the
principal dimensions varied during workload construction. The table
describes the collection principle rather than fixing a specific command
or tool to each class.

\begin{table}[t]
    \centering
    \caption{Network behaviors and workload variations used for NIC EM
    trace collection.}
    \label{tab:behavior_workloads}
    \footnotesize
    \setlength{\tabcolsep}{3.5pt}
    \renewcommand{\arraystretch}{1.08}
    \begin{tabular}{p{0.19\columnwidth} p{0.73\columnwidth}}
        \toprule
        \textbf{Behavior} & \textbf{Workload construction and variations} \\
        \midrule
        Normal &
        Active Blog browsing and content access; varied resources,
        navigation order, and request intervals. \\

        Flood &
        High-rate requests with varied rate, concurrency, and timing. \\

        PortScan &
        Port discovery with varied scan configuration, rate, and target
        ports. \\

        ServiceProbe &
        Service/version probing with varied probe configuration, timing,
        and target services. \\

        DirEnum &
        Directory enumeration with varied wordlists, request rates, and
        concurrency. \\

        PathDisc &
        Web-path discovery with varied path sets, request patterns, and
        concurrency. \\

        SQLi &
        SQL-injection probing with varied inputs, timing, and request
        patterns. \\

        XSS &
        Cross-site-scripting probing with varied inputs, timing, and
        request patterns. \\
        \bottomrule
    \end{tabular}
\end{table}

The threat-related workloads are not collected by suspending the normal
service and executing an attack in isolation. Legitimate requests
continue to reach the Blog service while threat-related activities are
performed. Consequently, the measured EM signal reflects aggregate NIC
activity under concurrent network processing. A trace labeled
\emph{PortScan}, for example, denotes an observation acquired while port
scanning is present in the monitored workload; it should not be
interpreted as a pure electromagnetic waveform generated exclusively by
the scanner. The same principle applies to the other threat classes.

After acquisition, each measurement used for recognition is associated
with its behavior label. We denote the resulting dataset as
\begin{equation}
    \mathcal{D}
    =
    \{(\mathbf{x}_i,y_i)\}_{i=1}^{N},
    \qquad
    \mathbf{x}_i
    =
    [x_i[0],\ldots,x_i[L_i-1]],
    \label{eq:trace_dataset}
\end{equation}
where \(y_i\in\mathcal{Y}\) is the network behavior associated with the
observation and \(L_i\) is the number of acquired samples. We do not
require all original acquisitions to share the same temporal extent;
this variability is handled in the representation stage described
next.

\subsection{Time--Frequency Representation}
\label{subsec:tf_representation}

\begin{figure*}[!ht]
    \centering
    \includegraphics[width=0.95\textwidth]{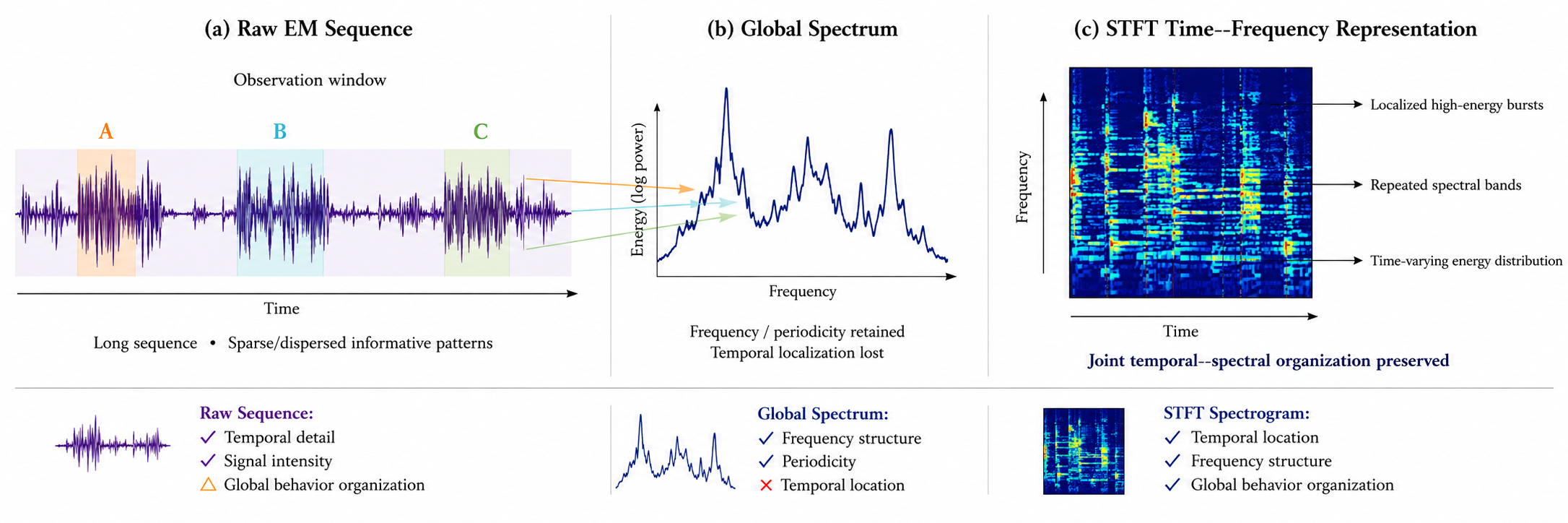}
    \caption{Comparison of raw-sequence, global-spectrum, and STFT representations.}
    \label{fig:representation_motivation}
\end{figure*}

The raw NIC EM observation preserves fine temporal detail, but
behavior-related information may be distributed across burst timing,
repetition, periodicity, and spectral-energy variation. A whole-trace
Fourier spectrum summarizes frequency content but removes temporal
localization. NICWhisper therefore uses a time--frequency
representation to retain both dimensions~\cite{allen1977stft,jin2023timefrequency}.

Figure~\ref{fig:representation_motivation} illustrates this distinction.
The raw sequence retains fine temporal detail, but informative structures
can remain dispersed over the observation. The global spectrum exposes
frequency and periodic information while collapsing the locations at
which these components occur. A time--frequency representation preserves
both aspects and makes localized bursts, repeated spectral structures,
and time-varying energy distributions directly observable in the same
representation.

NICWhisper applies only limited signal conditioning before
time--frequency transformation. We do not attempt to identify packet
boundaries or explicitly separate benign and threat-related components
from the measured trace. Non-finite samples, when present, are handled
deterministically, and the trace-wise DC component is removed:
\begin{equation}
    \tilde{x}_i[n]
    =
    x_i[n]
    -
    \frac{1}{L_i}
    \sum_{r=0}^{L_i-1}x_i[r].
    \label{eq:dc_removal}
\end{equation}
This removes the measurement offset while retaining relative amplitude
variation and temporal organization.

\textbf{Key Design 2: Preserving joint temporal--spectral
organization.}
For a conditioned observation \(\tilde{\mathbf{x}}_i\), NICWhisper
computes the short-time Fourier transform~\cite{allen1977stft}
\begin{equation}
    X_i(m,k)
    =
    \sum_{n=0}^{W-1}
    \tilde{x}_i[n+mH]\,
    w[n]\,
    e^{-j2\pi kn/K},
    \label{eq:stft}
\end{equation}
where \(w[n]\) is the analysis window, \(W\) is the window length,
\(H\) is the hop size, \(K\) is the FFT size, and \(m\) and \(k\)
index time frames and frequency bins, respectively. We use its
log-power representation
\begin{equation}
    S_i(m,k)
    =
    \log\!\left(1+\left|X_i(m,k)\right|^2\right),
    \label{eq:log_power}
\end{equation}
which reduces the dynamic range while retaining the organization of
spectral energy over time.

The STFT reorganizes the observation so that local frequency content is associated with its temporal position. Sustained
activity may therefore appear as persistent energy structure, repeated
processing as recurring spectral patterns, and intermittent probing as
temporally localized activity. These structures need not form
deterministic signatures; the representation simply exposes dimensions
along which behavior-related differences can be learned.

The time--frequency representation also provides a practical way to
handle acquisition traces with different lengths. For a trace with
\(L_i\) samples, the number of STFT frames depends on its duration, and
the resulting spectrogram may therefore have a different temporal
dimension from that of another trace. Rather than padding or truncating
a potentially very long raw waveform to a common number of samples,
NICWhisper maps the spectrogram to a fixed spatial resolution:
\begin{equation}
    \bar{\mathbf{S}}_i
    =
    \mathcal{R}(\mathbf{S}_i)
    \in
    \mathbb{R}^{F_0\times T_0},
    \label{eq:spectrogram_resize}
\end{equation}
where \(\mathcal{R}(\cdot)\) denotes the deterministic resizing
operation and \(F_0\times T_0\) is the common classifier input size.
This operation provides a consistent classifier input while retaining the coarse ordering of temporal and spectral structures.

All samples use the same STFT and resizing configuration. Feature
normalization is estimated from the training data and applied unchanged
to validation and test observations. Concrete values for the STFT
window, hop size, input resolution, and other implementation parameters
are reported in Section~\ref{sec:evaluation_setup}.

\subsection{Network Behavior Recognition}
\label{subsec:behavior_recognition}

NICWhisper finally learns the association between the time--frequency
representation and the corresponding network behavior. Because NICWhisper focuses on characterizing the physical observation channel, we use a compact conventional classifier for behavior recognition.

The classifier operates directly on
\(\bar{\mathbf{S}}_i\). It consists of a small sequence of standard
two-dimensional convolutional blocks~\cite{krizhevsky2012imagenet} followed by global feature
aggregation and a classification layer. The convolutional operations
capture local time--frequency structures, while the progressively
aggregated representation combines evidence distributed across the
spectrogram. Detailed layer dimensions and training hyperparameters are
given in Section~\ref{sec:evaluation_setup}.

The classifier is trained using the standard cross-entropy objective. For a previously unseen NIC EM observation, the signal undergoes exactly
the same conditioning, STFT computation, log-power transformation,
spatial resizing, and normalization as the training samples. The
classifier then predicts one of the behavior classes in
\(\mathcal{Y}\). No packet contents or host-side states are introduced
at this stage.

\section{Evaluation}
\label{sec:evaluation}

We evaluate NICWhisper to determine whether NIC electromagnetic
leakage can provide a meaningful physical observation of network
behaviors. Our evaluation characterizes the NIC EM observation channel along five dimensions. Specifically, we ask five questions.
RQ1 examines whether NIC EM measurements contain sufficient
information to distinguish benign activity and different threat
behaviors. RQ2 investigates what signal structure supports this
recognition capability. RQ3 studies whether behavior-related
information persists across execution variations. RQ4 characterizes
the sensitivity of the observation under controlled measurement
perturbations. Finally, RQ5 examines whether the learned
behavior-related information transfers across different NIC hardware.

\subsection{Experimental Setup}
\label{sec:evaluation_setup}

\textbf{Physical measurement platform.}
We use the same NIC EM acquisition platform throughout the evaluation.
The sensing target is an external USB~3.0 Gigabit Ethernet adapter
based on the Realtek RTL8153B controller. A magnetic near-field probe is
fixed on a three-axis precision positioning stage and placed near the
target NIC to maintain consistent sensing geometry across measurements.
The captured analog signal is amplified by a 30-dB low-noise amplifier
and digitized using a PicoScope 3000-series oscilloscope at
1~MS/s. The sensing and acquisition equipment remains external to the
monitored software environment.

The monitored network service remains active during collection, as
described in Section~\ref{subsec:trace_construction}. Normal samples
contain active legitimate network interactions rather than an idle NIC,
while threat-related samples are acquired in the presence of concurrent
legitimate activity. Detailed workload-generation tools and behavior-specific execution configurations are provided in Appendix~\ref{app:dataset_generation}.

\textbf{Dataset and split.}
The main source-NIC acquisition contains 4,652 valid one-second
EM traces across eight behavior classes. Because the original class
sizes differ, we construct a balanced evaluation pool by selecting
500 observations per class, yielding 4,000 traces in total.
Table~\ref{tab:data_split} summarizes the fixed training, validation,
and test partitions used in RQ1, RQ2, and RQ4. The selection seed
is fixed to 42. RQ3 and RQ5 follow their dedicated evaluation
protocols described in Sections~V-D and~V-F, respectively. The split is acquisition-session disjoint: all traces from the same execution session are assigned to a single partition.
\begin{table}[t]
    \centering
    \caption{Balanced source-NIC data split used in RQ1, RQ2, and RQ4.}
    \label{tab:data_split}
    \footnotesize
    \setlength{\tabcolsep}{6pt}
    \begin{tabular}{lccc}
    \rowcolor{tableheader}
    \toprule
    Partition & Per Class & Classes & Total \\
    \midrule
    Training   & 300 & 8 & 2,400 \\
    Validation & 50  & 8 & 400   \\
    Test       & 150 & 8 & 1,200 \\
    \bottomrule
    \end{tabular}
\end{table}

\textbf{Signal representation.}
Each evaluation trace contains \(10^6\) samples acquired at 1~MS/s,
corresponding to a one-second observation. We first replace isolated
non-finite samples with zero and remove the trace-wise DC component.
NICWhisper then computes a log-power STFT using a 2,048-sample Hann
window~\cite{harris1978windows}, a 2,048-point FFT, and a non-overlapping 2,048-sample hop.
The resulting one-sided representation contains 1,025 frequency bins
and 488 temporal frames over the 0--500~kHz frequency range.

The spectrogram is area-resampled to \(128\times128\) before
classification. We standardize the resulting representations using one
global mean and standard deviation estimated exclusively from the
training set. Validation and test samples reuse these statistics without
refitting.

\textbf{Recognizer and training.}
The classifier contains three convolutional stages with 32, 64, and
128 channels. The first two stages are followed by \(2\times2\)
max pooling, while the final stage is followed by adaptive global
average pooling and an eight-class linear prediction layer. The model
contains 94,152 trainable parameters.

We optimize cross-entropy loss using AdamW~\cite{loshchilov2017decoupled} with a learning rate of
\(10^{-3}\), weight decay \(10^{-4}\), and batch size 32. Training is
limited to 20 epochs, with early stopping after five epochs without
improvement in validation Macro-F1. The best validation checkpoint is
used for testing. Unless otherwise stated, every experiment is repeated
with five random seeds, \(0\)--\(4\), and results are reported as
mean~\(\pm\)~standard deviation. All experiments are executed on an
Intel Xeon w5-3423 CPU without GPU acceleration.

\textbf{Metrics.}
We report Accuracy together with macro-averaged Precision, Recall, and
F1-score. Macro-F1 is used as the primary multi-class metric because it
assigns equal importance to all eight behaviors. For RQ1, we
additionally report a pooled benign-versus-threat AUROC and analyze the
geometry of the learned representation. For robustness experiments,
the same trained models and test observations are retained so that
performance changes can be attributed to the evaluated perturbation.

\subsection{RQ1: Observation Feasibility}
\label{subsec:rq1}

\textbf{Can NIC electromagnetic leakage provide sufficient physical
information to distinguish benign activity and different network threat
behaviors?}

RQ1 evaluates the central hypothesis of NICWhisper: whether network
behaviors leave sufficiently informative physical evidence in the
electromagnetic emissions of a NIC to support recognition. We evaluate
the eight-class task using the fixed balanced test set and repeat the
experiment over five independent training runs.

Table~\ref{tab:rq1_perclass} reports the per-class and overall
recognition performance. NICWhisper achieves an average Accuracy of
\(80.53\%\) and a Macro-F1 of \(80.67\%\). Since the test set contains
the same number of observations for every behavior, the Macro-F1
provides a direct view of recognition quality across the complete
behavior set rather than being dominated by a particular class.

\begin{table*}[t]
    \centering
    \caption{Per-class and overall recognition performance (\%).}
    \label{tab:rq1_perclass}

    \footnotesize
    \renewcommand{\arraystretch}{1.14}
    \setlength{\tabcolsep}{4.2pt}

    \begin{tabularx}{\textwidth}{lYYYYYYYY}
        \toprule
        \rowcolor{tableheader}
        \textbf{Metric} &
        \textbf{Normal} &
        \textbf{Flood} &
        \textbf{PortScan} &
        \textbf{SvcProbe} &
        \textbf{DirEnum} &
        \textbf{PathDisc} &
        \textbf{SQLi} &
        \textbf{XSS} \\
        \midrule

        \textbf{F1-score} &
        $84.85 \pm 6.82$ &
        $90.51 \pm 1.85$ &
        $69.12 \pm 10.74$ &
        $79.33 \pm 15.13$ &
        $75.06 \pm 4.66$ &
        $70.98 \pm 6.01$ &
        $\mathbf{91.44 \pm 2.16}$ &
        $84.04 \pm 5.95$ \\

        Precision &
        $78.90 \pm 9.38$ &
        $93.14 \pm 4.73$ &
        $69.61 \pm 9.18$ &
        $86.27 \pm 13.47$ &
        $84.62 \pm 13.86$ &
        $64.99 \pm 12.81$ &
        $\mathbf{95.56 \pm 3.55}$ &
        $91.84 \pm 3.65$ \\

        Recall &
        $\mathbf{92.53 \pm 6.76}$ &
        $88.27 \pm 3.67$ &
        $71.47 \pm 20.06$ &
        $74.53 \pm 17.17$ &
        $69.47 \pm 9.70$ &
        $81.60 \pm 11.48$ &
        $87.87 \pm 4.43$ &
        $78.53 \pm 12.02$ \\
        \midrule

        \rowcolor{tableavg}
        \multicolumn{9}{l}{
            \textbf{Overall:}\quad
            Accuracy: $\mathbf{80.53 \pm 5.14}$ \qquad
            Macro-F1: $\mathbf{80.67 \pm 5.36}$ \qquad
            Macro-Precision: $83.12 \pm 5.39$ \qquad
            Macro-Recall: $80.53 \pm 5.14$
        } \\
        \bottomrule
    \end{tabularx}
\end{table*}

The results are not driven by a single easily recognizable behavior.
Flood and SQLi obtain F1-scores of \(90.51\%\) and \(91.44\%\),
respectively, while Normal and XSS also exceed \(84\%\). PortScan and
PathDisc are comparatively more challenging, with F1-scores of
\(69.12\%\) and \(70.98\%\). ServiceProbe and DirEnum lie between
these two groups. This variation is expected for a physical observation
channel: different behaviors can induce partially overlapping NIC
activity even though their high-level network semantics differ.

\begin{figure*}[t]
    \centering
    \includegraphics[width=\textwidth]{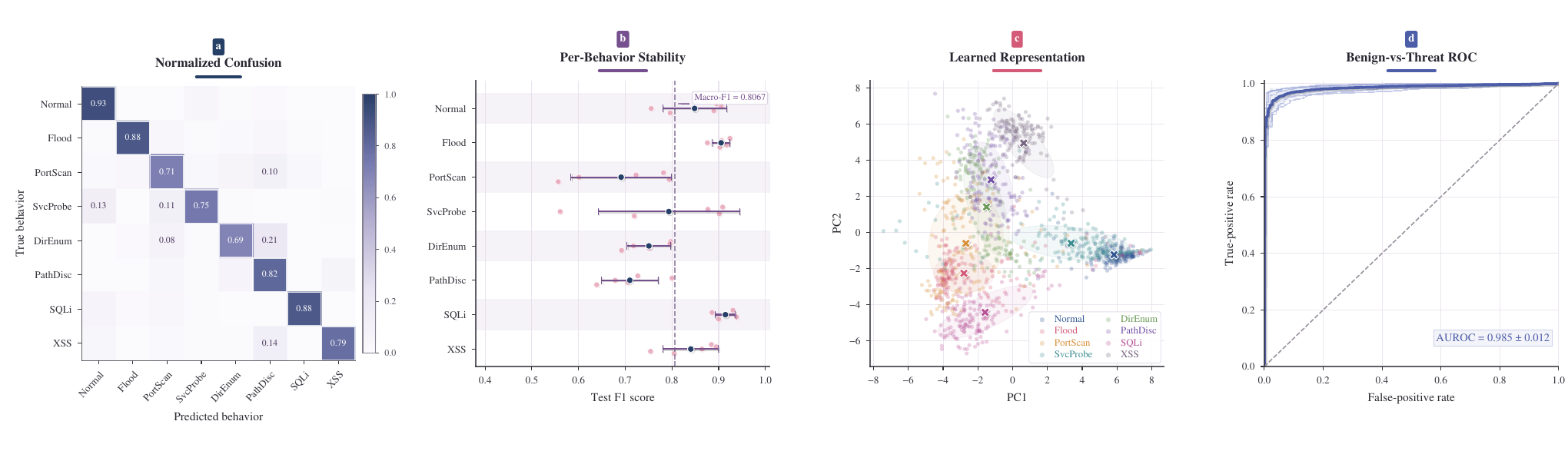}
    \caption{Analysis of NICWhisper's recognition capability in RQ1.
    (a) Row-normalized confusion matrix aggregated over five runs.
    (b) Per-behavior F1 distributions across the five runs, with
    mean and standard deviation; the dashed line denotes the overall
    Macro-F1.
    (c) PCA visualization of the 128-dimensional penultimate-layer
    representations for a representative run whose Macro-F1 is closest
    to the five-run mean.
    (d) Benign-versus-threat ROC curves obtained directly from the
    eight-class predictions using \(1-p(\mathrm{Normal})\) as the
    threat score.}
    \label{fig:rq1_analysis}
\end{figure*}

To examine the recognition behavior beyond aggregate metrics,
Fig.~\ref{fig:rq1_analysis} provides four complementary views of the
results.

Figure~\ref{fig:rq1_analysis}(a) shows that the dominant mass remains
on the diagonal of the normalized confusion matrix, confirming that
the measured channel contains information beyond a binary
active-versus-inactive distinction. The remaining errors are
concentrated among several behaviors that can induce similar
request-driven activity. In particular, DirEnum is frequently confused
with PathDisc, while ServiceProbe exhibits partial confusion with
Normal and PortScan. Such errors are consistent with the fact that the
EM observation reflects aggregate NIC processing rather than explicit
packet or application semantics.

Figure~\ref{fig:rq1_analysis}(b) further examines the result across
independent runs. Flood, SQLi, Normal, and XSS remain comparatively
stable, whereas PortScan and ServiceProbe exhibit larger run-to-run
variation. Nevertheless, every behavior retains recognition performance
substantially above the chance level of the eight-class task. The observed variation is more consistent with differences in class difficulty than with a complete loss of discriminative information for particular behaviors.

For each test observation, we extract the 128-dimensional feature vector
immediately before the final classification layer. As illustrated in
Fig.~\ref{fig:rq1_analysis}(c), the representations exhibit visible
class-dependent organization while retaining substantial overlap,
consistent with the errors observed in the confusion matrix. Across the
five runs, the mean within-class distance is
\(3.149\pm0.099\), whereas the mean distance between class centroids is
\(6.820\pm0.226\). Their ratio is \(2.166\pm0.049\), indicating
that between-class separation is approximately twice the average
within-class dispersion. The corresponding silhouette score is
\(0.178\pm0.029\). We do not interpret this representation as a set of
deterministic EM signatures; rather, it provides complementary evidence
that the measured traces contain learnable behavior-related structure.

Finally, we consider a coarser security question: whether the NIC EM
observation can distinguish active benign behavior from the presence of
any evaluated threat behavior. We do not train a separate binary model.
Instead, for each eight-class prediction, we define the threat score as
\(1-p(\mathrm{Normal})\), treating Normal as the negative class and all
seven threat behaviors as positive. As shown in
Fig.~\ref{fig:rq1_analysis}(d), this yields a mean AUROC of
\(0.985\pm0.012\) across five runs. The substantially higher
benign-versus-threat separability compared with the finer eight-class
task suggests that a considerable portion of the remaining errors occurs
among threat behaviors rather than between benign activity and threat
presence.

\subsection{RQ2: What Signal Structure Supports Behavior Recognition?}
\label{subsec:rq2}

RQ1 demonstrates that NIC electromagnetic measurements contain sufficient information to distinguish different network behaviors. However, recognition performance alone does not reveal how behavior-related information is represented in the physical signal. As NIC emissions reflect traffic-induced processing activity, this information may be distributed across signal magnitude, temporal dynamics, and spectral structure. RQ2 therefore examines how recognition changes when these signal properties are selectively suppressed or disrupted.

We consider three types of diagnostic controls. First, amplitude and
energy normalization reduce global signal-strength cues while retaining
most of the temporal organization of the observation. Second, temporal
and spectral interventions perturb one structural dimension while
leaving other information partially available. Finally, time--frequency
shuffling disrupts the joint spatial organization of the spectrogram.
We additionally include a random-label control as a sanity check. These
interventions are diagnostic analyses rather than models of physical
attacks against NICWhisper.

\begin{figure}[t]
    \centering
    \includegraphics[width=0.9\linewidth]{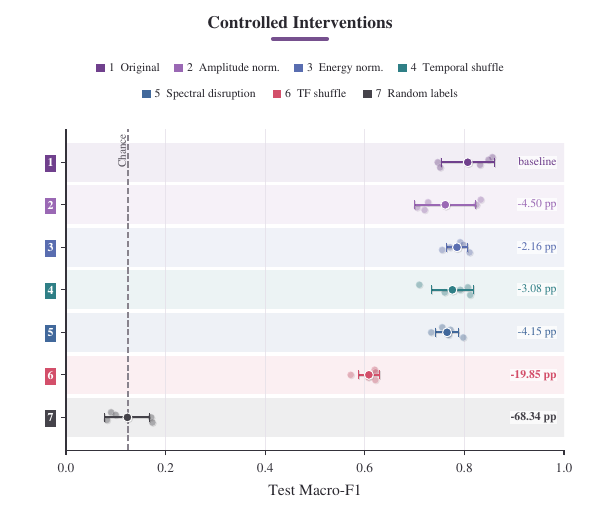}
    \caption{Recognition performance under the controlled interventions in RQ2. Error bars denote mean $\pm$ standard deviation over five runs.}
    \label{fig:rq2_interventions}
\end{figure}

Figure~\ref{fig:rq2_interventions} summarizes the recognition results
under these controls.

The original representation achieves a Macro-F1 of
\(80.67\%\pm5.36\%\). After amplitude normalization, performance
remains at \(76.17\%\pm6.15\%\), corresponding to a decrease of
4.50 percentage points. Energy normalization produces an even smaller
reduction of 2.16 points, retaining \(78.51\%\pm2.18\%\) Macro-F1.
Thus, absolute signal magnitude and aggregate energy contribute to
recognition, but neither can account for the majority of the
discriminative capability observed in RQ1.

Perturbing temporal or spectral organization individually produces a
similarly moderate effect. Temporal block shuffling yields
\(77.59\%\pm4.22\%\) Macro-F1, while spectral structure disruption
yields \(76.52\%\pm2.36\%\), corresponding to decreases of 3.08 and
4.15 percentage points, respectively. These results suggest that useful
information is not confined to only one of the two dimensions. When one
form of organization is weakened, other properties of the physical
observation remain available to the classifier.

\begin{figure}[!t]
    \centering
    \includegraphics[width=0.85\linewidth]{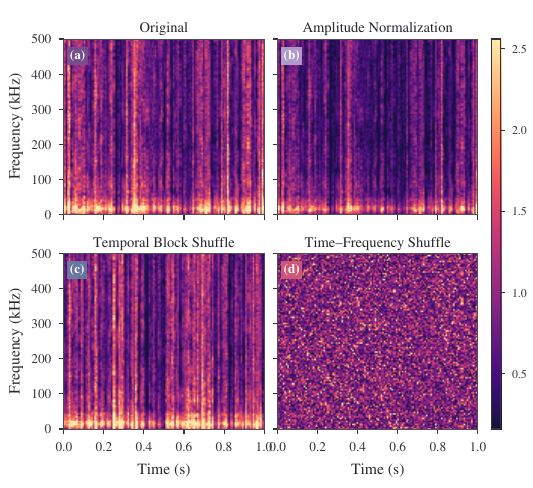}
    \caption{Effect of representative interventions on the same NIC EM
    observation. (a) Original time--frequency representation.
    (b) Amplitude normalization changes the global scale while retaining
    much of the temporal--spectral organization.
    (c) Temporal block shuffling changes the long-range temporal
    arrangement while preserving local structures.
    (d) Time--frequency shuffling strongly disrupts joint spatial
    organization.}
    \label{fig:rq2_examples}
\end{figure}

A substantially larger reduction occurs when the joint
time--frequency organization is disrupted. Time--frequency shuffling
reduces Macro-F1 to \(60.82\%\pm2.13\%\), a decrease of
19.85 percentage points relative to the original representation. This
drop is markedly larger than those caused by suppressing global
magnitude cues or perturbing temporal and spectral structure
individually. The result provides direct experimental support for the
representation choice in Section~\ref{subsec:tf_representation}:
recognition benefits from how NIC activity is jointly organized across
time and frequency rather than from a single global signal statistic.

The random-label control provides a final sanity check. When the input
observations remain unchanged but their training labels are randomized,
Macro-F1 falls to \(12.33\%\pm4.50\%\), close to the
\(12.5\%\) chance level of the eight-class problem. This confirms that
the performance observed in RQ1 depends on a meaningful association
between NIC EM observations and network-behavior labels rather than on
the training procedure or class distribution alone.

To make the structural interventions more interpretable,
Fig.~\ref{fig:rq2_examples} visualizes representative transformations
applied to the same NIC EM observation. 

As illustrated in Fig.~\ref{fig:rq2_examples}(a)--(b), amplitude
normalization leaves the locations and relative organization of many
time--frequency structures visually intact, consistent with its modest
effect on recognition performance. Temporal block shuffling preserves
substantial local spectral content but changes its arrangement along the
time axis. In contrast, time--frequency shuffling largely removes the
coherent neighborhood structure visible in the original representation.
The visual progression is therefore consistent with the quantitative
results: weakening simple global cues has a limited effect, whereas
destroying joint temporal--spectral organization produces a much larger
loss in recognition capability.

\subsection{RQ3: Attack-Execution Variation}
\label{subsec:rq3}

\textbf{Does behavior-related NIC EM information persist when the
concrete execution of the same network behavior changes?}

The experiments in RQ1 establish that NIC electromagnetic observations
are discriminative under the main evaluation setting. A remaining
question is whether this capability is tied to the concrete execution
profile used to generate each behavior. For example, a classifier might
associate a behavior with the traffic rhythm induced by one specific
tool or parameter configuration rather than with information that
persists across different realizations of the same network behavior.

RQ3 therefore varies two components of an execution while keeping the
behavior label unchanged. The first is the \emph{parameter
configuration}, which changes execution conditions such as rate,
timing, concurrency, scan configuration, or other behavior-specific
parameters. The second is the \emph{tool/implementation variant}, which
changes the implementation used to realize the same behavior. We denote
the three parameter configurations as P1--P3 and the three
tool/implementation variants as Tool-A--Tool-C. All protocols cover the
same eight behavior classes. Concrete mappings and configuration
details are provided in Appendix~\ref{app:dataset_generation}.

We evaluate two complementary protocols. The
\emph{mixed-variation} protocol contains all nine combinations of
Tool-A--Tool-C and P1--P3. Training, validation, and test partitions are acquisition-session disjoint, while each partition contains execution diversity.
This setting evaluates whether NICWhisper can learn behavior information
when tool and parameter variation are already present in the
observation pool.

The \emph{held-out-variation} protocol is more restrictive. One entire
parameter configuration or tool variant is excluded from training and
appears only during testing. For parameter variation, we evaluate
P1+P2$\rightarrow$P3 and P2+P3$\rightarrow$P1. For tool variation, we
evaluate Tool-A+Tool-B$\rightarrow$Tool-C and
Tool-A+Tool-C$\rightarrow$Tool-B. Thus, the evaluated execution
identity is unavailable to the classifier during training.

Table~\ref{tab:rq3_variation} summarizes the protocols, execution
coverage, data volume, and recognition results.

\begin{table*}[t]
    \centering
    \caption{Recognition under mixed and held-out execution variations in RQ3 (\%).}
    \label{tab:rq3_variation}

    \footnotesize
    \setlength{\tabcolsep}{5.0pt}
    \renewcommand{\arraystretch}{1.13}

    \resizebox{\textwidth}{!}{%
    \begin{tabular}{
        l
        l
        l
        l
        c
        c
        c
        c
        >{\columncolor{tableavg}}c
    }
        \toprule

        \rowcolor{tableheader}
        \textbf{Protocol} &
        \textbf{Factor} &
        \textbf{Train Vars.} &
        \textbf{Test Var.} &
        \textbf{\#Scen.} &
        \textbf{Obs. (Tr/V/Te)} &
        \textbf{Accuracy (\%)} &
        \textbf{Macro-F1 (\%)} &
        \textbf{Avg. (\%)} \\
        \midrule

        \textbf{Mixed} &
        Tool+Param. &
        A/B/C $\times$ P1/P2/P3 &
        Mixed pool &
        9 &
        2400/400/1200 &
        $78.30 \pm 4.62$ &
        $78.06 \pm 4.88$ &
        -- \\

        \midrule

        \multirow{2}{*}{\textbf{Param. hold-out}} &
        \multirow{2}{*}{Parameter} &
        P1 + P2 &
        P3 &
        3 &
        4800/800/1200 &
        $75.00 \pm 5.66$ &
        $73.77 \pm 6.24$ &
        \textbf{(Acc.) $71.47 \pm 4.45$} \\

        &
        &
        P2 + P3 &
        P1 &
        3 &
        4800/800/1200 &
        $67.95 \pm 5.86$ &
        $67.51 \pm 5.93$ &
        \textbf{(F1) $70.64 \pm 5.04$} \\

        \midrule

        \multirow{2}{*}{\textbf{Tool hold-out}} &
        \multirow{2}{*}{Tool} &
        Tool-A + Tool-B &
        Tool-C &
        3 &
        4800/800/1200 &
        $86.93 \pm 1.83$ &
        $86.88 \pm 1.84$ &
        \textbf{(Acc.) $85.07 \pm 1.54$} \\

        &
        &
        Tool-A + Tool-C &
        Tool-B &
        3 &
        4800/800/1200 &
        $83.20 \pm 2.71$ &
        $83.16 \pm 2.86$ &
        \textbf{(F1) $85.02 \pm 1.61$} \\

        \bottomrule
    \end{tabular}%
    }
\end{table*}

The mixed-variation protocol first examines recognition when execution
diversity is present throughout the dataset. It contains nine distinct
Tool$\times$Parameter identities and 2,400/400/1,200
training/validation/test observations. NICWhisper achieves
\(78.30\%\pm4.62\%\) Accuracy and
\(78.06\%\pm4.88\%\) Macro-F1. The result shows that combining
multiple concrete executions within each behavior class does not remove
the discriminative information available in the NIC EM channel.

We next withhold an entire parameter configuration from training. When P1 and P2 are available during training and P3 is
reserved exclusively for testing, NICWhisper achieves
\(73.77\%\pm6.24\%\) Macro-F1. In the reverse configuration,
training with P2 and P3 and testing on the unseen P1 configuration
produces \(67.51\%\pm5.93\%\) Macro-F1. The aggregate result across
the two directions is \(70.64\%\pm5.04\%\).

The difference between the two parameter-held-out directions is itself
informative. It indicates that execution parameters can induce
substantial changes in the measured physical distribution even when the
high-level behavior remains unchanged. Nevertheless, both tests retain
considerable multi-class recognition capability despite complete
exclusion of the evaluated parameter configuration from the training
protocol.

We next change the tool or implementation while preserving the behavior
label. Training on Tool-A and Tool-B and evaluating exclusively on
Tool-C yields \(86.88\%\pm1.84\%\) Macro-F1. Training on Tool-A and
Tool-C and evaluating on the unseen Tool-B yields
\(83.16\%\pm2.86\%\). Across the two directions, the average Macro-F1
is \(85.02\%\pm1.61\%\).

Importantly, the higher absolute values observed in the tool-held-out
experiments should \emph{not} be interpreted as evidence that an unseen
tool improves recognition. The RQ1, mixed-variation, and held-out
protocols do not share identical training constructions or test
distributions. The underlying base split remains 2,400/400/1,200
training/validation/test observations. In each held-out experiment,
however, every base training and validation observation is instantiated
under the two available training-side variations, producing 4,800
training and 800 validation observations, whereas each base test
observation is instantiated only under the held-out variation, leaving
1,200 test observations. Moreover, individual held-out variants can
differ in intrinsic classification difficulty. Absolute scores across
these protocols are therefore not a controlled measure of performance
gain or loss.

Instead, the relevant result is that withholding the evaluated
tool/implementation does not cause recognition to collapse. The
classifier continues to distinguish the eight behavior classes even
when the implementation used at test time is entirely absent from its
training variations. Thus, despite changes in the concrete
implementation used to realize a behavior, the behavior-related
information captured in the NIC EM observations remains recognizable
under previously unseen tool/implementation variants.

\subsection{RQ4: Sensitivity to Controlled Signal Perturbations}
\label{subsec:rq4}

\textbf{How sensitive is NICWhisper to controlled perturbations of the
measured EM observation?}

\begin{figure*}[!t]
    \centering
    \includegraphics[width=\textwidth]
    {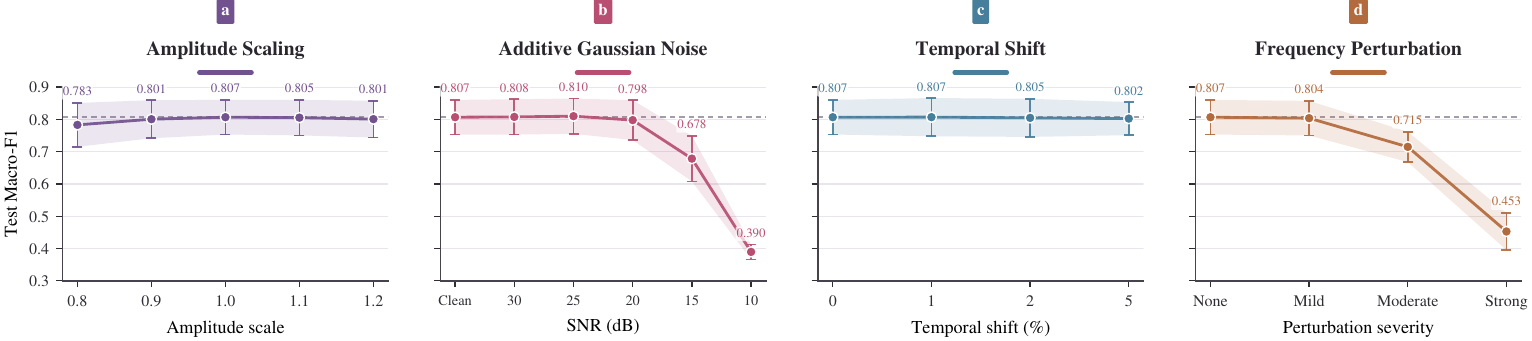}
    \caption{Sensitivity of NICWhisper to controlled signal perturbations without retraining. Points report mean Macro-F1 over five runs and error
    bars denote standard deviation. The dashed line marks the original
    RQ1 performance. (a) Amplitude scaling. (b) Additive white Gaussian
    noise at different SNR levels. (c) Temporal shifts relative to the
    observation duration. (d) Increasing frequency-domain perturbation.}
    \label{fig:rq4_robustness}
\end{figure*}

The preceding experiments use EM traces acquired under a fixed
measurement configuration. Since NICWhisper relies on a physical observation channel, its measurements can vary in signal magnitude, background noise, temporal alignment, and spectral response~\cite{sehatbakhsh2020remote,ayoub2025phasesca}. We characterize these factors individually by perturbing one signal property at a time while keeping the acquisition configuration fixed.

We reuse the five models trained in RQ1 and evaluate them on the same
1,200-trace test set without retraining or adaptation. Four controlled
perturbations are considered: amplitude scaling, additive white Gaussian
noise, temporal shifting, and frequency-domain perturbation.
Figure~\ref{fig:rq4_robustness} reports the resulting Macro-F1 scores.
The dashed line marks the original RQ1 Macro-F1 of \(80.67\%\).

Figure~\ref{fig:rq4_robustness}(a) first examines changes in signal
magnitude. Across scaling factors from \(0.8\times\) to \(1.2\times\),
Macro-F1 ranges from \(78.29\%\) to \(80.67\%\). Performance remains
close to the original setting for \(0.9\times\)--\(1.2\times\), while
the \(0.8\times\) condition produces a somewhat larger but still
moderate decrease. These results indicate that the recognizer is not
critically dependent on one exact signal amplitude, although magnitude
variation is not completely inconsequential.

The additive-noise experiment in
Fig.~\ref{fig:rq4_robustness}(b) reveals a clearer
severity-dependent transition. Macro-F1 remains close to the original
result at 30~dB (\(80.76\%\)), 25~dB (\(80.99\%\)), and
20~dB (\(79.77\%\)) SNR. The small differences among these conditions
fall within the run-to-run variation and should not be interpreted as a
performance improvement caused by noise. At 15~dB, however, Macro-F1
drops to \(67.84\%\), and at 10~dB it further decreases to
\(38.97\%\). Thus, the system tolerates moderate additive noise in this
controlled setting, but recognition degrades sharply once noise becomes
sufficiently strong.

The nonlinear decrease is consistent with the perturbation magnitude
used in the experiment. Noise is scaled from each trace's raw
mean-square power, with the noise RMS equal to the trace RMS divided by
\(10^{\mathrm{SNR}/20}\). Consequently, reducing SNR from 20~dB to
10~dB increases the noise RMS by approximately \(3.16\times\) and the
noise power by \(10\times\). The intermediate 15~dB result shows that
the degradation is not an isolated failure at 10~dB, but rather the
continuation of a transition that begins as broadband contamination
becomes strong enough to obscure part of the time--frequency structure.

Temporal misalignment has substantially less impact. As shown in
Fig.~\ref{fig:rq4_robustness}(c), shifting the observation by
\(1\%\), \(2\%\), and \(5\%\) of its duration yields Macro-F1 scores
of \(80.69\%\), \(80.47\%\), and \(80.23\%\), respectively,
compared with \(80.67\%\) without shifting. NICWhisper therefore does
not require the relevant activity to occur at one precisely fixed
absolute location within the observation window.

Finally, Fig.~\ref{fig:rq4_robustness}(d) evaluates increasing
frequency-domain perturbation. A mild perturbation retains
\(80.37\%\) Macro-F1, whereas moderate perturbation reduces performance
to \(71.49\%\). Under the strongest evaluated perturbation, Macro-F1
falls further to \(45.29\%\). This trend is consistent with the RQ2
observation that structured spectral information contributes to
recognition: small spectral changes can be tolerated, whereas stronger
distortion progressively removes information available to the
time--frequency representation.

Taken together, the four experiments show that measurement sensitivity
is not uniform. Moderate amplitude variation, small temporal
misalignment, and moderate additive noise produce limited degradation,
whereas sufficiently strong broadband noise or spectral distortion
substantially reduces recognition performance. These results therefore
identify an operating region rather than robustness to arbitrary
measurement degradation.

\subsection{RQ5: Cross-Device Transfer}
\label{subsec:rq5}

\textbf{Does behavior-related NIC EM information transfer to unseen NIC
hardware?}

Since electromagnetic emissions depend on the underlying hardware~\cite{laor2022drawnapart,peng2025adaptive}, we
evaluate whether the behavior-related information learned from one NIC
remains useful after the monitored device changes. We use the Realtek
RTL8153B as the source NIC and evaluate four previously unseen target
NICs: Realtek RTL8156BG, ASIX AX88179A, Microchip LAN7800, and WCH
CH9200. The target devices cover both same-vendor and cross-vendor hardware. The acquisition chain and signal-processing configuration are kept unchanged.

\begin{table*}[ht]
    \centering
    \caption{Cross-device transfer from the source RTL8153B NIC to
    four previously unseen NICs. Direct transfer uses no target-device
    training data, whereas 10\% fine-tuning uses a small target-domain
    training subset. Results are mean $\pm$ standard deviation over
    five runs (\%).}
    \label{tab:rq5_crossdevice}

    \footnotesize
    \setlength{\tabcolsep}{9.0pt}
    \renewcommand{\arraystretch}{1.14}

    \begin{tabular}{lllcccc}
        \toprule

        \rowcolor{tableheader}
        & & &
        \multicolumn{2}{c}{\textbf{Direct Cross-Device}} &
        \multicolumn{2}{c}{\textbf{10\% Fine-tuning}} \\[-0.15ex]

        \arrayrulecolor{black!55}
        \cline{4-5}\cline{6-7}
        \arrayrulecolor{black}

        \rowcolor{tableheader}
        \multirow{-2}{*}{\textbf{Target NIC}} &
        \multirow{-2}{*}{\textbf{Vendor}} &
        \multirow{-2}{*}{\textbf{Relation to Source}} &
        \textbf{Accuracy} &
        \textbf{Macro-F1} &
        \textbf{Accuracy} &
        \textbf{Macro-F1} \\
        \midrule

        RTL8156BG &
        Realtek &
        Same vendor, different model &
        $73.52 \pm 5.18$ &
        $72.67 \pm 5.36$ &
        $81.28 \pm 0.72$ &
        $\mathbf{80.47 \pm 0.84}$ \\

        AX88179A &
        ASIX &
        Cross-vendor, different model &
        $70.24 \pm 3.32$ &
        $69.48 \pm 3.41$ &
        $81.47 \pm 0.83$ &
        $\mathbf{80.23 \pm 0.91}$ \\

        LAN7800 &
        Microchip &
        Cross-vendor, different model &
        $69.36 \pm 4.07$ &
        $68.42 \pm 4.25$ &
        $80.63 \pm 0.94$ &
        $\mathbf{79.51 \pm 1.02}$ \\

        CH9200 &
        WCH &
        Cross-vendor, different model &
        $68.17 \pm 3.46$ &
        $67.35 \pm 3.53$ &
        $78.42 \pm 1.13$ &
        $\mathbf{77.37 \pm 1.21}$ \\

        \bottomrule
    \end{tabular}
\end{table*}

We consider two transfer settings. In \emph{direct cross-device}
evaluation, the model is trained exclusively on the RTL8153B source
device and applied to the target NIC without using any target-device
training or validation data. This setting measures how much
behavior-related information transfers directly across the hardware
change. We additionally consider a lightweight adaptation setting in
which only 10\% of the target-domain training data is used to fine-tune
the source model, while the target test set remains unchanged. All
results are averaged over five independent random seeds.

Table~\ref{tab:rq5_crossdevice} shows that changing the NIC introduces
a clear distribution shift, but does not eliminate the behavior-related
information captured by NICWhisper. Without any target-device training
data, Macro-F1 ranges from \(67.35\%\) to \(72.67\%\) across the four
target NICs, with an average of \(69.48\%\). The highest direct-transfer
performance is obtained on the RTL8156BG at \(72.67\%\), while the
three cross-vendor NICs achieve \(69.48\%\), \(68.42\%\), and
\(67.35\%\), respectively.

These results are substantially above the \(12.5\%\) chance level of
the eight-class task, indicating that part of the behavior-related
time--frequency structure learned on the source NIC remains useful on
previously unseen hardware. At the same time, the decrease relative to
the source-device setting confirms that the physical observation is not
device invariant. NIC-specific circuitry and electromagnetic coupling
affect the measured distribution and therefore remain an important
source of domain shift.

A small amount of target-device adaptation substantially reduces this
gap. With only 10\% target-domain fine-tuning data, Macro-F1 increases
to \(80.47\%\), \(80.23\%\), \(79.51\%\), and \(77.37\%\) on the four
target NICs, respectively. Averaged across devices, Macro-F1 improves
from \(69.48\%\) under direct transfer to \(79.40\%\) after fine-tuning.
The improvement is consistent across all four target devices, ranging
from 7.80 to 11.09 percentage points.

These results indicate that cross-device measurements contain both transferable behavior structure and device-specific variation. The transferable component supports direct recognition on unseen NICs, while a small amount of target data substantially reduces the device-induced distribution shift.

\section{Related Work}
\label{sec:related_work}

\textbf{Network and host-based threat monitoring.}
Existing network threat detection primarily relies on traffic-visible
or host-visible evidence. Network-based approaches analyze packets,
flows, protocol characteristics, statistical patterns, or encrypted
traffic structures to identify malicious activities
~\cite{alkasassbeh2023intrusion,khraisat2019survey,wei2023xnids,
han2024ecnet,shen2021accurate,fu2024flow}. Host-based and
provenance-based systems instead characterize suspicious execution from
processes, system events, audit records, and causal relationships
~\cite{zipperle2022provenance,jia2024magic,cheng2024kairos,
zhang2025tapas,wang2026dinspector,aly2025ocr,chen2022apt}. At the communication-interface level, OASIS embeds intrusion
detection directly into BLE controllers
~\cite{cayre2024oasis}. These approaches provide substantially richer semantic
information than physical measurements, but their acquisition depends
on visibility into the communication path or the monitored software
stack.

\textbf{Electromagnetic leakage and high-level activity inference.}
Electromagnetic side-channel analysis was initially studied mainly for
recovering cryptographic and low-level computational information
~\cite{quisquater2001electromagnetic,picek2023sok}. Subsequent work,
however, has shown that unintended emissions can preserve substantially
higher-level information about device operation. EM Eye demonstrates
that electromagnetic leakage from embedded cameras can reveal visual
content~\cite{long2024emeye}, while other studies infer sensitive
information from GPUs, fingerprint sensors, and mobile-device
components~\cite{zhan2022graphics,ni2023recovering,
chen2024eavesdropping}. EM radiation has also been exploited for
detecting and localizing hidden electronic devices, including GPS
trackers and hidden cameras~\cite{li2024gpsbuster,zhang2024eye}.
Together, these studies show that externally observable EM emissions can
retain structured information associated with internal device activity,
rather than merely exposing simple variations in signal magnitude.

Recent work further demonstrates that physical leakage can expose
complex workload structure. Maia et al. recover neural-network topology
and execution characteristics through GPU magnetic
emissions~\cite{maia2022hear}, while ModelSpy exploits
architecture-dependent electromagnetic modulation for long-range model
architecture inference~\cite{xiao2026modelspy}. Related studies have
investigated neural-network extraction and GPU side-channel leakage
through power and electromagnetic observations
~\cite{gao2024deeptheft,horvath2024sok,horvath2025barracuda}.
These works are particularly relevant because they demonstrate that
physical emissions may preserve temporal and structural characteristics
of a workload even when the underlying computation is complex. 

\textbf{Physical side channels as defensive sensors.}
Physical side channels have also been explored as external security
sensors. WattsUpDoc uses power consumption to detect malware on embedded
medical devices without modifying the monitored software
~\cite{clark2013wattsupdoc}, while DeepPower exploits power
side-channel measurements for non-intrusive IoT malware
detection~\cite{ding2020deeppower}. Electromagnetic sensing has enabled
similar defensive mechanisms. EMMA uses externally measured EM signals
for hardware/software attestation~\cite{sehatbakhsh2019emma}, and IDEA
detects intrusions in critical embedded and cyber-physical systems by
identifying deviations from expected electromagnetic execution
patterns~\cite{khan2021idea}. REMOTE further studies robust external malware detection using EM
observations across embedded and cyber-physical platforms
~\cite{sehatbakhsh2020remote}, while Pham et al. exploit EM side-channel
measurements to classify malware and obfuscated malware executions on
IoT devices~\cite{pham2021obfuscation}. Han et al. examine the effectiveness and
security limits of physical side-channel control-flow
monitoring~\cite{han2022hiding}. EM and power measurements have also
been applied to hardware-Trojan and anomalous-device
detection~\cite{sun2021electromagnetic,zhang2017hardware,
liao2024hardware}. These studies establish an important precedent for
using unintended physical emissions not only as leakage sources, but
also as independent observations for security monitoring. Closer to communication-interface hardware, BlueScream demonstrates
that protocol-driven digital activity in BLE devices can induce
externally observable electromagnetic leakage
~\cite{ayoub2024bluescream}, while early NIST research showed that
WLAN cards exhibit measurable electromagnetic signatures that can
differentiate network devices~\cite{remley2005wlan}. 

\section{Discussion}
\label{sec:discussion}

\subsection{Implications}
\label{subsec:implications}

NICWhisper provides a physical observation channel that differs from conventional packet- and host-based monitoring by capturing unintended NIC electromagnetic emissions outside the monitored software stack. This makes it possible to obtain complementary behavior-level evidence without accessing packet contents or host-generated telemetry. The measured emissions reflect aggregate NIC activity and are influenced by traffic characteristics such as timing, rate, concurrency, and burst patterns. Our experiments show that these physical patterns remain useful across different execution settings, moderate measurement perturbations, and multiple NICs, although their distribution still depends on sensing conditions and hardware. Recognition degrades under severe signal distortion or substantial hardware changes, indicating that sensing quality and device-specific calibration are important considerations for practical use.

\subsection{Limitations}
\label{subsec:limitations}

NICWhisper has several limitations that bound the conclusions of the
current study. First, our evaluation covers five NICs under a controlled
acquisition setup with fixed sensing hardware and controlled probe
placement. Although the cross-device experiments demonstrate that
behavior-related information remains partially transferable across NIC
models and vendors, our goal in this work is to establish the
feasibility and characterize the properties of NIC EM leakage as a
network-security observation channel, rather than to claim a
deployment-ready detection system. A practical deployment may involve
additional variations in probes, acquisition devices, host platforms,
probe placement, and physical environments, which remain to be
systematically evaluated.

Second, the controlled-perturbation experiments consider controlled signal-level
perturbations, including amplitude variation, additive noise, temporal
shifting, and spectral distortion. These experiments characterize the
operating range of the observation channel but cannot fully reproduce
the coupled variations encountered in real environments, such as probe
displacement, nearby electronics, environmental interference, and
long-term measurement drift. Evaluating NICWhisper under realistic and
continuously operating deployment conditions, including sensing
calibration and maintenance, is therefore an important direction for
future work.

Finally, the current evaluation considers a predefined set of behavior
families and active benign workloads. Extending NIC EM sensing to
broader benign activities and previously unseen network behaviors will
require additional data collection and open-world recognition
mechanisms. Moreover, NICWhisper provides behavior-level physical
evidence from aggregate NIC activity; fine-grained packet contents,
protocol semantics, or process-level reconstruction are beyond the
scope of this study.

\section{Conclusion}
\label{sec:conclusion}

This work investigates NIC electromagnetic leakage as an external
physical observation source for network threat-behavior recognition.
NICWhisper transforms raw EM measurements into time--frequency
representations and uses a lightweight classifier to distinguish benign
and threat-related behaviors. Experiments show that aggregate NIC processing produces behavior-related physical structure shaped by traffic dynamics and extending beyond simple signal strength; this structure remains informative across execution variations and moderate measurement perturbations and partially transfers to unseen NIC hardware. Although severe signal degradation and hardware shifts expose
clear limitations, the results establish NIC EM leakage as a promising
complementary physical observation surface for network security
monitoring.
Code and dataset information are available at: \url{https://anonymous.4open.science/r/NICWhisper-C838/}.

\section*{Acknowledgments}

Generative AI tools, including OpenAI ChatGPT, were used for language
refinement and illustrative figure creation. All AI-assisted content was manually reviewed by the authors.

\section*{Ethics Considerations}

All evaluated network activities were conducted in a controlled
experimental environment using research-operated systems and
infrastructure. No unauthorized third-party systems were scanned or
attacked. NICWhisper records electromagnetic measurements of NIC
activity and does not require packet contents, credentials, or personal
user information. The study is intended to investigate NIC
electromagnetic leakage as a defensive physical observation channel.
\bibliographystyle{IEEEtran} 
\bibliography{ref}

\clearpage

\appendices

\section{Detailed Results of Controlled Interventions}
\label{app:rq2_results}

Table~\ref{tab:rq2_detailed} reports the complete numerical results of
the controlled-intervention experiments in RQ2. All conditions follow
the same evaluation protocol as the original NICWhisper configuration
and are repeated over five independent runs. In addition to Macro-F1,
which is used as the primary metric in the main text, we report Accuracy
and the absolute Macro-F1 change relative to the original
representation.

\begin{table}[ht]
    \centering
    \caption{Detailed results of the controlled interventions used in
    RQ2. Values are mean $\pm$ standard deviation over five runs.
    $\Delta$F1 denotes the change in Macro-F1 relative to the original
    representation.}
    \label{tab:rq2_detailed}
    \footnotesize
    \setlength{\tabcolsep}{3.5pt}
    \renewcommand{\arraystretch}{1.10}
    \begin{tabular}{lccc}
        \toprule
        \rowcolor{tableheader}
        \textbf{Condition} &
        \textbf{Accuracy (\%)} &
        \textbf{Macro-F1 (\%)} &
        \textbf{$\Delta$F1 (pp)} \\
        \midrule

        Original &
        $80.53 \pm 5.14$ &
        $80.67 \pm 5.36$ &
        -- \\

        Amplitude Norm. &
        $76.15 \pm 6.05$ &
        $76.17 \pm 6.15$ &
        $-4.50$ \\

        Energy Norm. &
        $78.68 \pm 1.96$ &
        $78.51 \pm 2.18$ &
        $-2.16$ \\

        Temporal Shuffle &
        $77.98 \pm 4.11$ &
        $77.59 \pm 4.22$ &
        $-3.08$ \\

        Spectral Disruption &
        $76.83 \pm 2.02$ &
        $76.52 \pm 2.36$ &
        $-4.15$ \\

        Time--Frequency Shuffle &
        $61.85 \pm 1.67$ &
        $60.82 \pm 2.13$ &
        $\mathbf{-19.85}$ \\

        Random Labels &
        $15.38 \pm 7.06$ &
        $12.33 \pm 4.50$ &
        $-68.34$ \\

        \bottomrule
    \end{tabular}
\end{table}

The detailed results reinforce the comparison reported in the main
text. Normalizing global amplitude or energy reduces Macro-F1 by only
2.16--4.50 percentage points. Perturbing temporal or spectral
organization individually produces reductions of a similar magnitude,
at 3.08 and 4.15 points, respectively. In contrast, destroying the
joint time--frequency organization results in a 19.85-point decrease.

These controls should be interpreted comparatively rather than as
independent estimates of feature importance. For example, the decrease
under amplitude normalization indicates that absolute signal magnitude
does contain useful information; the remaining \(76.17\%\) Macro-F1
shows only that this cue is insufficient to explain the original
recognition result. Likewise, temporal and spectral structure each
contribute to classification while leaving complementary information
available when disrupted individually. The larger degradation under
joint time--frequency shuffling is therefore the key observation: the
recognizer benefits from structured information distributed across both
dimensions.

The random-label result provides a separate sanity control rather than a
signal intervention. Its \(12.33\%\) Macro-F1 is close to the
eight-class chance level and confirms that meaningful correspondence
between the measured observations and their behavior labels is required
for successful recognition.

\section{Controlled Signal Perturbations}
\label{app:robustness}

Table~\ref{tab:app_rq4} summarizes the controlled signal perturbations
used in RQ4. All perturbations are applied to the raw test observations
before the standard NICWhisper preprocessing and STFT transformation.
The five models trained in RQ1 are kept fixed throughout these
experiments.

For amplitude scaling, each trace is multiplied by
$\alpha\in\{0.8,0.9,1.0,1.1,1.2\}$.
For additive Gaussian noise, zero-mean white Gaussian noise is added at
SNR levels of 30, 25, 20, 15, and 10~dB. Given the RMS value $r_x$ of
a trace, the noise standard deviation is
\[
\sigma_n=\frac{r_x}{10^{\mathrm{SNR}/20}}.
\]
Temporal perturbation shifts the observation by 1\%, 2\%, or 5\% of its
duration, in addition to the unshifted condition.

For frequency perturbation, we apply deterministic band-selective
attenuation to the raw observation in the Fourier domain. Let
$Z[k]=\mathcal{F}_{r}\{x\}[k]$ denote the rFFT coefficients of a raw
trace $x$, where $k$ indexes the positive-frequency range from DC to the
Nyquist frequency. For a selected frequency interval
$\mathcal{B}=[f_l,f_h)$ and attenuation factor $\gamma$, the perturbed
spectrum is defined as
\[
\widetilde{Z}[k]=
\begin{cases}
\gamma Z[k], & k\in\mathcal{B},\\
Z[k], & \text{otherwise}.
\end{cases}
\]
The perturbed time-domain observation is then reconstructed as
\[
\widetilde{x}=\mathcal{F}_{r}^{-1}\{\widetilde{Z}\},
\]
after which the standard NICWhisper preprocessing, STFT, log-power
transformation, and spatial resizing are applied.

Mild, Moderate, and Strong perturbations attenuate the
20--25\%, 18--32\%, and 15--40\% intervals of the positive-frequency
range using amplitude factors of 0.75, 0.50, and 0.25, respectively.
With the 1-MHz sampling rate, these intervals correspond to
100--125~kHz, 90--160~kHz, and 75--200~kHz. The power within the
attenuated band is therefore scaled by $\gamma^2$. All frequency
perturbations are deterministic: the same frequency interval and
attenuation factor are applied to every trace for a given severity
level, without randomness.

\begin{table}[ht]
\centering
\caption{Controlled signal perturbations used in RQ4.}
\label{tab:app_rq4}
\footnotesize
\setlength{\tabcolsep}{4.5pt}
\renewcommand{\arraystretch}{1.12}
\begin{tabular}{p{0.30\linewidth}p{0.62\linewidth}}
\toprule
\rowcolor{tableheader}
\textbf{Perturbation} & \textbf{Evaluated settings} \\
\midrule

Amplitude scaling &
$0.8,\,0.9,\,1.0,\,1.1,\,1.2$ \\

Gaussian noise &
Clean, 30, 25, 20, 15, 10~dB \\

Temporal shift &
$0\%,\,1\%,\,2\%,\,5\%$ \\

Frequency perturbation &
None;
Mild: 20--25\%, $\gamma=0.75$;
Moderate: 18--32\%, $\gamma=0.50$;
Strong: 15--40\%, $\gamma=0.25$ \\

\bottomrule
\end{tabular}
\end{table}
\section{Dataset Generation Protocol: Tool Variants and Parameter Configurations}
\label{app:dataset_generation}

\begin{table*}[h]
\centering
\caption{Tool variants and behavior-specific configuration dimensions used
for threat-workload construction in RQ3. P1--P3 correspond to different
combinations of the listed parameters.}
\label{tab:tool_variants}
\scriptsize
\setlength{\tabcolsep}{4.0pt}
\renewcommand{\arraystretch}{1.15}

\begin{tabular}{
p{0.080\textwidth}
p{0.235\textwidth}
p{0.410\textwidth}
p{0.205\textwidth}}
\toprule
\rowcolor{tableheader}
\textbf{Behavior} &
\textbf{Tool Variant} &
\textbf{Varied Configuration Dimensions} &
\textbf{Representative Command} \\
\midrule

\multirow{3}{*}{\textbf{Flood}}
& Tool-A: hping3
& \multirow{3}{0.410\textwidth}{
Packet transmission mode and rate control; destination-port selection;
source-address and packet-field randomization; burst timing and request
scheduling.
}
& \texttt{hping3 -S --flood -p 80 <IP>} \\

& Tool-B: Metasploit SYN Flood
&
& \path{auxiliary/dos/tcp/synflood} \\

& Tool-C: Custom Scapy Generator
&
& \texttt{syn\_flood.py --rate 1000 --rand-src} \\
\midrule

\multirow{3}{*}{\textbf{PortScan}}
& Tool-A: Nmap SYN Scan
& \multirow{3}{0.410\textwidth}{
Scan mode; host-discovery configuration (e.g., default discovery or
\texttt{-Pn}); timing template; target-port selection; probe retry/timeout
settings; and scan-rate control.
}
& \texttt{nmap -sS -Pn -T4 <IP>} \\

& Tool-B: Masscan
&
& \texttt{masscan -p1-65535 --rate=1000 <IP>} \\

& Tool-C: Metasploit TCP Portscan
&
& \path{auxiliary/scanner/portscan/tcp} \\
\midrule

\multirow{3}{*}{\textbf{ServiceProbe}}
& Tool-A: Nmap Version Detection
& \multirow{3}{0.410\textwidth}{
Target-service selection; service/version probe mode; probe intensity;
timeout and retry settings; service-discovery options; and auxiliary
probe/script selection where supported.
}
& \texttt{nmap -sV -T4 <IP>} \\

& Tool-B: Amap
&
& \texttt{amap -B <IP> <ports>} \\

& Tool-C: Custom Banner Grabber
&
& \texttt{banner\_grab.py --ports 1-1000} \\
\midrule

\multirow{3}{*}{\textbf{DirEnum}}
& Tool-A: Gobuster
& \multirow{3}{0.410\textwidth}{
Wordlist selection; worker/thread count; recursion behavior; traversal
depth; extension set; request timeout; and response/status filtering.
}
& \texttt{gobuster dir -u <URL> -w <list> -t 50} \\

& Tool-B: Dirb
&
& \texttt{dirb <URL> <wordlist>} \\

& Tool-C: Feroxbuster
&
& \texttt{feroxbuster -u <URL> -w <list> -t 50} \\
\midrule

\multirow{3}{*}{\textbf{PathDisc}}
& Tool-A: Dirsearch
& \multirow{3}{0.410\textwidth}{
Path-dictionary selection; HTTP request method; extension set; request
concurrency; recursion behavior; request scheduling; and header/options
used during path probing.
}
& \texttt{dirsearch -u <URL>} \\

& Tool-B: WFuzz
&
& \texttt{wfuzz -z file,<list> <URL>/FUZZ} \\

& Tool-C: Custom Requests Fuzzer
&
& \texttt{path\_disc.py --wordlist <list>} \\
\midrule

\multirow{3}{*}{\textbf{SQLi}}
& Tool-A: sqlmap
& \multirow{3}{0.410\textwidth}{
Target-parameter selection; request method; injection technique and payload
set; probing depth; level/risk configuration; crawling behavior; and
DBMS-oriented fingerprinting options.
}
& \texttt{sqlmap -u "<URL>?id=1" --batch} \\

& Tool-B: Metasploit SQLi Scanner
&
& \path{auxiliary/scanner/http/sql_injection} \\

& Tool-C: Custom Blind-SQLi Fuzzer
&
& \texttt{sqli\_fuzz.py --payloads <list>} \\
\midrule

\multirow{3}{*}{\textbf{XSS}}
& Tool-A: XSSer
& \multirow{3}{0.410\textwidth}{
Target-parameter coverage; GET/POST request selection; payload family and
encoding; crawling behavior; injection context; DOM-related checks; and
response-verification strategy.
}
& \texttt{xsser -url "<URL>" --auto} \\

& Tool-B: XSStrike
&
& \texttt{xsstrike.py -u "<URL>?q="} \\

& Tool-C: Custom XSS Fuzzer
&
& \texttt{xss\_fuzz.py --payloads <list>} \\

\bottomrule
\end{tabular}
\end{table*}
This appendix details the tool implementations and execution configurations
used to construct the threat workloads for the execution-variation experiments
in Section~\ref{subsec:rq3}. For each behavior, three distinct tools or
implementations (Tool-A--Tool-C) are combined with three predefined
behavior-specific parameter profiles (P1--P3). These profiles correspond
to different combinations of multiple execution parameters rather than
ordered intensity levels. Table~\ref{tab:tool_variants} summarizes the
tools and the configuration dimensions varied when constructing these
profiles. All threat workloads are executed in the presence of concurrent
benign background traffic.

For each behavior, the three tool variants are combined with the three
parameter profiles, resulting in nine Tool$\times$Parameter execution
identities. P1--P3 use different combinations of the behavior-specific
configuration dimensions summarized above. Additional randomized
micro-variations, including timing jitter, target or path selection,
payload ordering, and request scheduling, are introduced during collection
to avoid repeatedly reproducing an identical execution trace.

\end{document}